\documentclass[epjST]{svjour}

\usepackage{graphics}

\usepackage{amsmath,amssymb}

\begin{document}

\title{Symmetry breaking induced tipping to aging}

\author{I. Gowthaman\inst{1} \and V. K. Chandrasekar\inst{1}  
\and D. V. Senthilkumar\inst{2} \and M. Lakshmanan \inst{3}}

\institute{ Centre for
  Nonlinear Science and Engineering, School of Electrical and
  Electronics Engineering, SASTRA Deemed University, Thanjavur -
  613401, Tamilnadu, India \and School of Physics, Indian Institute of Science Education
  and Research, Thiruvananthapuram - 695551, India \and Department of  Nonlinear Dynamics, School of Physics, Bharathidasan University, Tiruchirappalli 620 024, India}

\abstract{
The competing effect of  heterogeneity and symmetry breaking coupling on the emerging dynamics in a system of $N$ globally coupled Stuart-Landau oscillators
is investigated.  Increasing the heterogeneity, using the standard deviation of the Hopf bifurcation parameter, favors the macroscopic oscillatory state for 
low values of the symmetry breaking coupling  and inhomogeneous steady state for larger values of the coupling.  There is also a transition, 
{\it tipping}, to homogeneous steady state (aging state)  from the macroscopic oscillatory state. The limiting factor in the diffusive coupling favors the macroscopic oscillatory state
even in the presence of a large fraction of inactive oscillators in the network thereby increasing the robustness of the network.  The globally coupled oscillators are reduced
to a system of two evolution equations for the macroscopic order parameters, corresponding to the mean-field and the shape parameter, using the self-consistent 
field approach.  The bifurcation diagrams obtained from the mean-field variables elucidate  various bifurcation scenarios responsible for  the dynamical transitions
observed in  $N$ globally coupled Stuart-Landau oscillators.  In particular, tipping to the aging state  is found to occur via the Hopf  and pitchfork bifurcations
illustrating the phenomenon of  {\it bifurcation induced tipping}. Analytical stability (critical) curves of these bifurcations, deduced from the mean-field variables,
are found to fairly well agree with the  simulation results.
}

\maketitle

\section{Introduction}
\label{intro}

Coupled oscillator network constitutes an excellent framework to understand a plethora of complex collective behavior.  The phenomenon of {\it aging} is 
such an intriguing collective behavior,  introduced to understand the effect of local deterioration/failures  on the macroscopic  dynamical state.  Originally, it was introduced
by Daido et al ~\cite{hdkn2004}, by increasing the  fraction of inactive oscillators in an ensemble of active oscillators to find out the threshold at which the macroscopic
oscillation ceases to exist. Such a threshold has also been identified as a measure of the dynamical robustness of the coupled networks, which
has actually been  rigorously deduced ranging  from regular to complex network topologies~\cite{hd2009,hd2011,whxz2014,gtkm2014,hd2008,btds2014,hdnk2007,tgmk2012,mktg2011}.  
Initially, {\it aging}  phenomenon has been reported in globally coupled oscillator networks~\cite{hdkn2004}, but later extended to  locally coupled oscillator networks by
 introducing local disorders~\cite{hd2009,hd2011,whxz2014}, heterogeneous oscillator networks with various topologies~\cite{gtkm2014,hd2008},
and also by introducing time-delay in the coupling~\cite{btds2014}. It has also been extended to complex networks~\cite{hdnk2007,tgmk2012} and multilayered networks~\cite{mktg2011}.
Interestingly, a desynchronization horn has been associated with the onset of the aging~\cite{hdnk2007}.

Recent activities on aging are centered at enhancing the robustness of the macroscopic oscillatory state despite the presence of a large fraction of inactive oscillators. 
For instance,  several methods such as  introducing auxiliary systems in the coupling~\cite{sksm2018},  introducing a simple limiting factor that limits the diffusive interaction~\cite{ylwz2016}, incorporating
large variance of the asymmetry parameter~\cite{sksm2018}, employing adaptive coupling  such as low-pass filters~\cite{usks2020}, introducing  asymmetry coupling among the populations~\cite{sksm2018},  and allowing normal and uniform random errors in the Hopf bifurcation parameters~\cite{ssnm2017}, etc.  have been 
shown to increase the robustness of the macroscopic oscillatory state over a  large range of coupling strength  and even shown to prevent the onset of  aging
in several scenarios despite the presence of a large fraction of the inactive oscillators.
 
Further,  the onset of inhomogeneous (asymmetric) dynamical states in  coupled oscillator networks has been known to arise largely due to symmetry breaking coupling, which breaks
the rotational symmetry of the entire system of equations.  In particular, the distinct 
oscillation quenching states, namely  amplitude and oscillation deaths, chimera death and so on, have been shown to be induced by breaking the rotational symmetry of the
coupled oscillator networks.  Notably, the dynamical transition from amplitude death to oscillation death via Turing bifurcation has been shown to get mediated by the 
spontaneous symmetry breaking  phenomenon~\cite{gsap2012,akev2013,ismk2015}.  Further, the loss of rotational symmetry also leads to breaking of
either translational or permutation symmetry of the coupled oscillator networks.  Recent  research reports unveiled that breaking the prevailing permutation/translational symmetry
of the coupled oscillators facilitates the emergence of diverse  macroscopic dynamical states such as chimera, solitary state, clustering, 
etc.~\cite{alev2013,azis2013,azmk2014,tb2015,kpvkc2015,kpvkc2016,kssk2017,ksvkc2018}.

Despite the growing body of literature on the phenomenon of {\it aging}, the  effect of symmetry breaking coupling on the same and in particular the role of
the competing interaction between the symmetry breaking coupling and the heterogeneity of the coupled oscillator networks on the aging has not yet been
explored in  great detail to the best of our knowledge.  In particular,  the nature of bifurcation that leads to the transition from the macroscopic oscillatory state
to the macroscopic quiescent state, namely {\it tipping} to the aging state,  of the coupled oscillator networks has not yet been identified.  In this work, 
using the coupled oscillator network comprising of active and inactive oscillators, we unravel the competing interaction between the symmetry  breaking
coupling and the degree of heterogeneity in understanding the dynamical robustness of the considered network. In particular, we find that increasing the standard deviation
of the Hopf bifurcation parameter, that is increasing the heterogeneity  of the network,  favors the macroscopic oscillatory state and  inhomogeneous dynamical states over
a large range of the parameter space. Further, the nature of the dynamical transition largely depends on the system frequency leading to the state dependent aging.
Limiting the diffusive interaction using the simple limiting factor favors macroscopic oscillatory state even in the presence of a large fraction of inactive oscillators
thereby increasing the robustness of the network. 
We deduce the amplitude of $N$ globally coupled oscillator network in terms of the global order parameters using  the mean-field approximation. The global order parameters
quantify the macroscopic oscillatory state. The stability conditions for the onset of  the aging state from the macroscopic oscillatory state and the 
inhomogeneous steady state are deduced from the evolution equations of the global order parameters, which exactly mimics the dynamical transitions 
obtained numerically from the $N$ globally coupled oscillator network. 
The bifurcation diagrams, obtained using XPPAUT~\cite{be2002},  from the evolution equations of the global order parameters also clearly elucidate the dynamical transitions.
Specifically,  we show that the onset of tipping to aging state occurs via the Hopf and inverse pitch-fork bifurcations thereby illustrating bifurcation induced tipping to the aging state.

The organization of the paper is as follows. In Sec. 2, we introduce a heterogeneous network  of $N$ globally coupled Stuart-Landau limit-cycle oscillators
with symmetry breaking coupling.   We discuss bifurcation induced tipping to the aging state using two-parameter
bifurcation diagrams in Sec. 3.  We deduce the evolution equations  corresponding to the  mean-field global order parameters from the system of 
$N$ globally coupled Stuart-Landau limit-cycle oscillators under the self-consistent field approach in Sec. 4 and  unravel the bifurcations mediating the
tipping to the aging state from  the evolution equations.  The effect of  the limiting factor 
on the observed dynamics is also discussed in Sec. 4.  Finally, we summarize our results and provide conclusion in Sec. 5.

\section{Model}
We consider a heterogeneous network of $N$ globally coupled paradigmatic model of Stuart-Landau oscillators with symmetry breaking coupling represented
 by the set of dynamical equations,
\begin{eqnarray}
\dot{z}_j &\,=(a_j+i\omega-|z_j|^2)z_j+\frac{k}{N}\sum_{l=1}^{N} Re\left({z_l}-\alpha {z_j}\right), \qquad i=1,2,\ldots, N
\label{model1}
\end{eqnarray}
where $z_j$ represents the position of the $j^{th}$ oscillator in the complex plane.   In Eq.~(\ref{model1}), $\omega$  is the intrinsic frequency, 
$a_j$ is the parameter specifying the distance of the $j^{th}$  oscillator from  the Hopf bifurcation value, which is drawn from a distribution with mean $a$ 
and variance $\sigma^2$,  $k$ is the  coupling strength  and {\bf $\alpha$ is the limiting factor~\cite{wzdv2015,dgtb2015}, which controls the rate of diffusion 
when $0\le\alpha\le 1$. The actual diffusive coupling is retrieved for $\alpha=1$ and direct feedback coupling for  $\alpha=0$.}
The coupling among the real variables  only results in  breaking the rotational symmetry $z_j\rightarrow z_je^{i\theta}$ of the globally coupled network.
The Stuart-Landau oscillator
represents the normal form of the Hopf bifurcation and hence a large collection of nonlinear oscillators exhibiting the Hopf bifurcation can be approximated as the 
Stuart-Landau oscillators. Hence,  an ensemble of  Stuart-Landau oscillators  serves as an excellent framework to unravel and understand a plethora of macroscopic dynamical
states that can be displayed by several other nonlinear oscillators.  In addition, Stuart-Landau oscillators can also open up  the feasibility of rigorous analytical treatment
in several circumstances.  Stuart-Landau oscillators exhibit limit-cycle oscillation for  $a_j>0$ and stable homogeneous steady state at the origin for $a_j<0$. 
In the following section, we will  discuss the dynamical transitions and spread of the dynamical states in the two-parameter phase diagrams  by
directly integrating Eq.~(\ref{model1}) numerically.

\begin{figure}
\centering
\resizebox{1.0\columnwidth}{!}{ \includegraphics{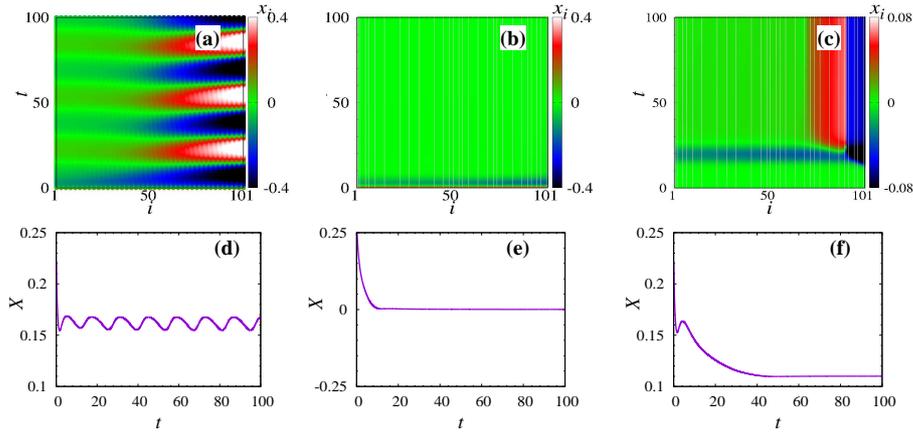} }
%\resizebox{0.85\columnwidth}{!}{ \includegraphics{Fig1b.eps} }
%\includegraphics[width=1.30\columnwidth]{ST.eps}
\caption{(color online) The spatiotemporal  behavior and the order parameter $X$, respectively, for $\omega=0.50$. (a, d) Limit cycle oscillations for $k=0.30$, (b, e)  homogeneous steady state (aging) for 
$k=1.5$ and  (c, f) inhomogeneous steady state (oscillation death) for $k=4.00$.  The mean of the distribution of the Hopf bifurcation parameters  is chosen as $a=-0.50$ with the 
standard deviation $\sigma=0.60$.}
	\label{fig1} 
\end{figure}
\section{Bifurcation induced tipping to aging: Simulation results}
The evolution equations corresponding to $N=100$ globally coupled Stuart-Landau oscillators are solved using the 
Runge-Kutta 4th order integration scheme with the time step $h=0.01$.  The limiting factor is taken as  $\alpha=1$. The Hopf bifurcation parameter $a_j$ is uniformly 
distributed with  $\left<a_j\right>=a=-0.5$ and standard deviation $\sigma=0.7$ so that a large fraction of oscillators are in their stable homogeneous steady state.  
The spatiotemporal plot and the order parameter $X=Re(Z)$, where $Z=\vert\frac{1}{N} \sum_{j=1}^{N} z_j\vert$,
as a function of time are depicted in Figs.~\ref{fig1} for $\omega=0.5$ and for different values of the coupling strength $k$.  The globally coupled Stuart-Landau oscillators
exhibit limit-cycle oscillations for $k=0.3$ despite the presence of a large fraction of inactive oscillators (see Fig.~\ref{fig1}(a)).   The  oscillatory nature of the order parameter $X$ 
 in Fig.~\ref{fig1}(d)   corroborates the limit-cycle oscillations for $k=0.3$. Further increase in the coupling strength results in tipping to the homogeneous steady state  (HSS), 
which is the aging state, as depicted in Fig.~\ref{fig1}(b)  for $k=1.5$, while the corresponding order parameter reaches  the null value after initial transients
 confirming the homogeneous steady state  at the origin.
However, the oscillators populate the branches of the inhomogeneous steady state (IHSS)  randomly  for $k=4$ as depicted in  Fig.~\ref{fig1}(c)   
and the corresponding order parameter acquires a finite value as illustrated in  Fig.~\ref{fig1}(f).
{\bf Note that the order parameter $X$ in Fig.~1(f) 
indeed corresponds to OD and not to nontrivial AD~\cite{dgtb2014}.}

\begin{figure*}[ht]
	\centering
	\hspace{-0.1cm}
	\resizebox{1.0\columnwidth}{!}{ \includegraphics{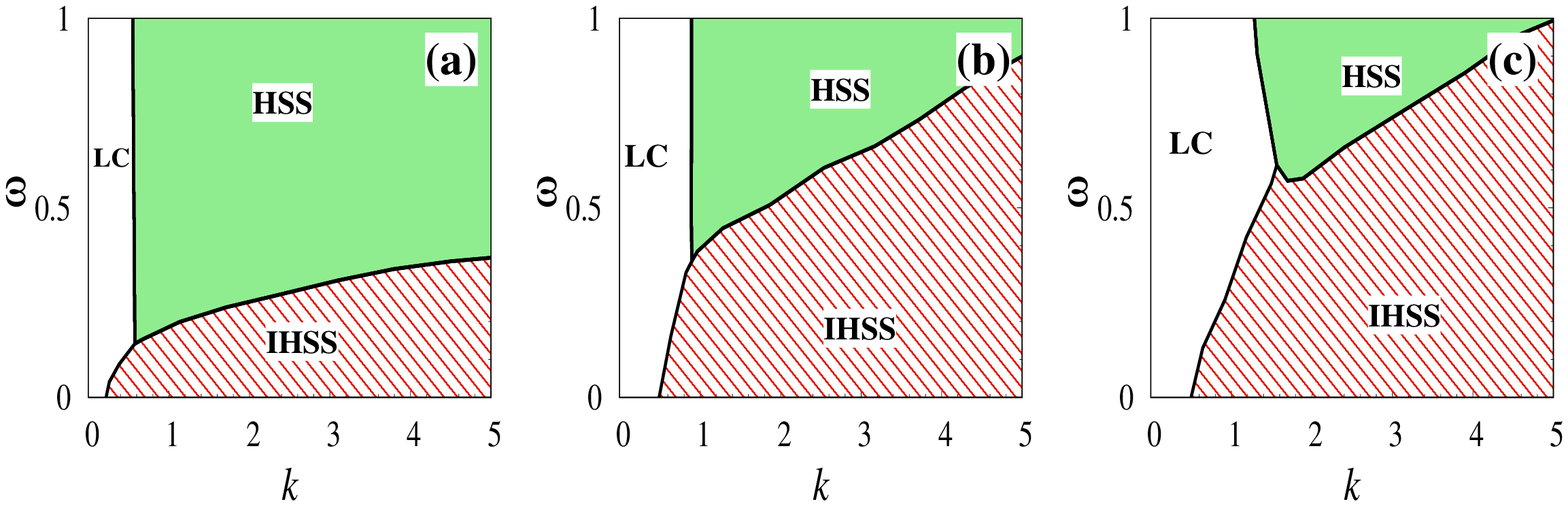} }
	\caption{(color online) Two-parameter phase diagrams in the  $(k, \omega)$  parameter space for different standard deviation $\sigma$ and  for $a=-0.5$.
		(a)  $\sigma=0.6$, (b) $\sigma=0.7$, and  (c) $\sigma=0.8$. }
	\label{fig2} 
\end{figure*}
The two-parameter phase diagrams as a function of $k$ and $\omega$ are depicted  in Fig.~\ref{fig2} for a global perspective of the dynamical transitions for different 
values of the standard deviation  with $a=-0.5$.  The parameter space with limit-cycle oscillations,  inhomogeneous steady state (oscillation death) and 
homogeneous steady state (aging) are indicated  by LC (unshaded region), 
IHSS (region with diagonal lines) and HSS (shaded region), respectively.   For $\sigma=0.6$, there is a transition from limit-cycle oscillations to IHSS as a function of the 
coupling strength $k$ for low values of the natural frequency $\omega\in[0,0.14)$ (see Fig.~\ref{fig2}(a)).  There is  also a transition from LC to IHSS via HSS in a
 rather narrow range of $\omega\in[0.14,0.37)$ as a function of $k$. 
Finally, there is a direct transition (tipping) from limit-cycle oscillations to aging state for $\omega\ge0.37$.    Such a frequency dependent tipping to aging is also known as state dependent
aging~\cite{igks2020}. Similar dynamical transitions are depicted in  Figs.~\ref{fig2}(b) and  ~\ref{fig2}(c) for $\sigma=0.7$ and $0.8$, respectively. 
It is to be noted that as the fraction of active
oscillators are increased for large values of the standard deviation $\sigma$, the spread of the aging state decreases while that of the limit-cycle oscillations and  IHSS increases. 
These results corroborate  the fact that  increasing  the value of standard deviation favors the macroscopic oscillatory state in a large range of the coupling strength $k$, thereby increasing
the robustness of the coupled network.  Furthermore, the threshold (minimum)  value of the natural frequency for the tipping to the aging also increases appreciably with $\sigma$. 
Due to the large heterogeneity for increasing $\sigma$, the spread of IHSS also increases to a large extent for large values of the coupling strength.
\begin{figure*}[ht]
	\centering
	\hspace{-0.1cm}
	\resizebox{1.0\columnwidth}{!}{ \includegraphics{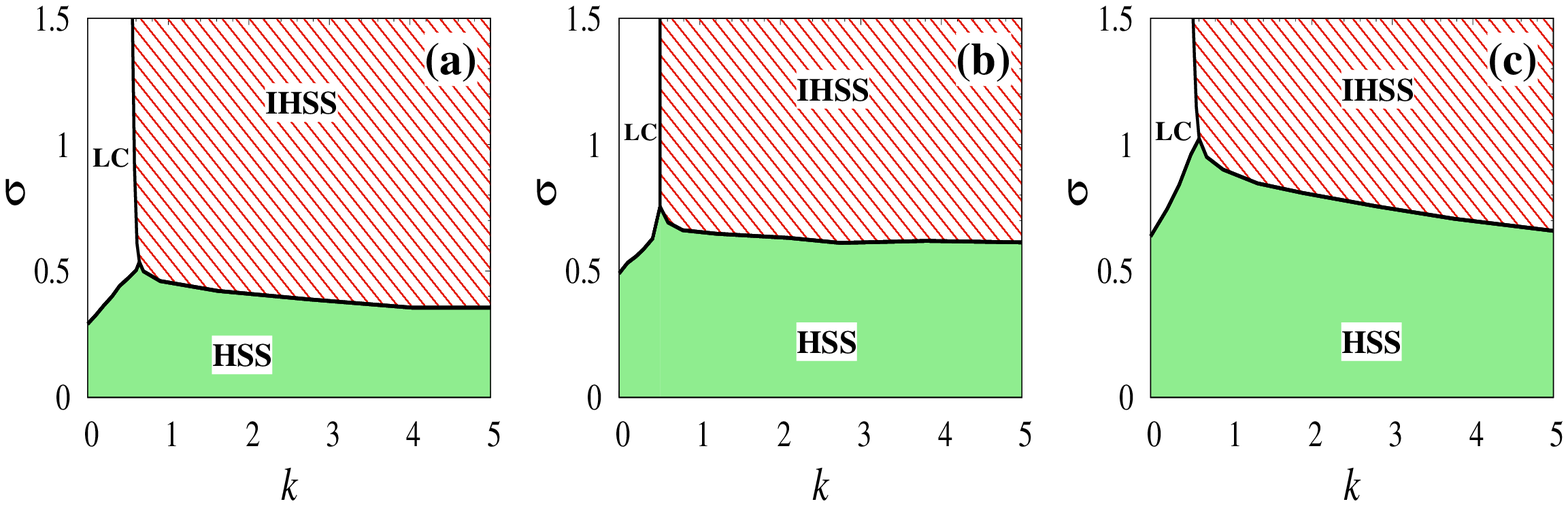} }
	\caption{(color online) Two-parameter phase diagrams in the  $(k, \sigma)$  parameter space for   different mean of the Hopf bifurcation parameter for  $\omega = 0.25$.
	(a) $a=-0.30$ (b) $a=-0.50$, and  (c) $a=-0.70$.}
	\label{fig3} 
\end{figure*}

Two-parameter phase diagrams in the $(k, \sigma)$ parameter space are depicted in Fig.~\ref{fig3} for different mean  values  of the Hopf bifurcation parameters $a=\left<a_j\right>$
for $\omega=0.25$. Only  the aging state is observed for $\sigma<0.3$ for $a=-0.3$  in the entire explored range of the coupling strength $k$ (see Fig.~\ref{fig3}(a)) because 
for $\sigma<0.3$,  the  value of the Hopf bifurcation parameter of all  the oscillators are less than zero attributing to the stable homogeneous steady state of the oscillators.  Similar scenario
prevails in Figs.~\ref{fig3}(b) and ~\ref{fig3}(c)  for $a=-0.5$ and $-0.7$, respectively.  For negative values of $a$, when the value of $\sigma$ 
increases beyond $\vert a\vert$, there exist some oscillators with
$a_j>0$ resulting in the macroscopic oscillatory state for small values of the coupling strength $k$.  However, for even slightly larger values of the symmetry breaking coupling $k$,
the heterogeneous nature of the oscillators plays a dominant role in facilitating the inhomogeneous steady state in a large range of $k$ and $\sigma$.  Nevertheless, there is
a transition from  macroscopic oscillatory state to IHSS via the aging state in a narrow range of $\sigma>\vert a\vert$.

\section{Mean-field approximation}
The system of globally coupled Stuart-Landu oscillators, Eq.~(\ref{model1}),  cannot be solved exactly. However,  it can be reduced to a system of two evolution equations for the
macroscopic order parameters, corresponding to the mean-field and the shape parameters, within the self-consistent field approach under the strong coupling limit~\cite{sdm2002,gmp2016}.
To be specific,  we adopt a mean-field approach proposed by Monte and Ovidio~\cite{sdm2002}  under a narrow distribution of the Hopf bifurcation parameter and in the 
strong coupling limit,  which is based on a  moment expansion and a closure assumption~\cite{sdm2002}.
Let $Z =\left< z_j \right>$ denotes the mean field $\frac{1}{N} \sum_{j=1}^{N} z_j$. 
As we can see $Z$ is the centroid of the $N$ oscillators, we can represent the position of the oscillators as  the distance from the centroid, $z_j = Z + \varepsilon_j$, where $\varepsilon_j$ is the distance of  the $j^{th}$ oscillator from the centroid $Z$. Since, $Z =\left< z_j \right>$, therefore $\left<\varepsilon_j \right>= 0$.
%which means the 1st moment of the distance of all the oscillators from their centroid, i.e. $\varepsilon_j$ , about origin is 0; considering the distance follows cartesian sign rule. 
By definition, $<a_j>= a$. It is also necessary to account for the dispersion of the $z_j$  about $Z$ by introducing the shape parameter,  denoted by $W =\left< a_j \varepsilon_j \right>$.
Now,  it is straightforward to arrive at a class of macroscopic variables, corresponding to the higher orders, as the equations of motion of the centroid. Nevertheless, many macroscopic properties
can be deduced even from the evolution equation of the macroscopic variables corresponding to the first order approximation~\cite{sdm2002,gmp2016}.  However,
we have also verified the emergence of  similar results for the second order approximation in the current investigation.

The evolution equations  corresponding to the centroid $Z$ and the shape parameter $W$ can be deduced, restricting to the first order approximation, 
under the narrow distribution of the Hopf bifurcation parameter $a_j$ and in the strong coupling limit as
\begin{eqnarray}
\dot{Z} &\,=(a+i\omega+(k/2)(1-\alpha)-|Z|^2)Z+(k/2)(1-\alpha)Z^*+W,
\label{mean1}
\end{eqnarray}
which has the same functional form as that of individual uncoupled elements in Eq.~(\ref{model1}), except  for the term $W$. Now, the time derivative of 
the shape parameter $W$ can be written as
\begin{eqnarray}
\dot{W} &\,=\left< a_j \dot{z}_j\right>-\left< a_j \right>\dot{Z},
\label{shape}
\end{eqnarray}
where, neglecting higher order terms, $\left\langle a_j \dot{z}_j \right\rangle$ can be deduced as
\begin{align}
\left\langle a_j \dot{z}_j \right\rangle&\,=\left\langle a_j^2 \right\rangle Z+\left\langle a_j^2 \epsilon_j\right\rangle +i\omega Z\left\langle a_j\right\rangle +i\omega \left\langle \epsilon_j a_j\right\rangle -|Z|^2 Z \left\langle a_j\right\rangle-2|Z|^2\left\langle \epsilon_j a_j\right\rangle-Z^2\left\langle \epsilon_j^* a_j\right\rangle  \nonumber\\ &\,+ 
 \frac{k}{2}(1-\alpha) Z\left\langle a_j\right\rangle-\frac{k}{2}\alpha\left\langle \epsilon_j a_j\right\rangle  + \frac{k}{2}(1-\alpha) Z^*\left\langle a_j\right\rangle-\frac{k}{2}\alpha\left\langle \epsilon_j^* a_j\right\rangle.
\label{az}
\end{align}

Substituting Eqs.~(\ref{az}) and ~(\ref{mean1})  in the evolution equation for the shape parameter (\ref{shape}), the latter can be rewritten,
neglecting $\left\langle (a_j-a)^2\epsilon_j\right\rangle$ due to the narrow distribution of the Hopf bifurcation parameter $a_j$, as
\begin{eqnarray}
\dot{W} &\,=\sigma^2 Z+(a-(\alpha k/2)-2|Z|^2+i\omega)W-(Z^2-(\alpha k/2))W^*,
\label{mean2}
\end{eqnarray}
where $\sigma^2 =\left\langle a_j^2 \right\rangle-a^2$. Equations~(\ref{mean1}) and ~(\ref{mean2}) govern the dynamics of the oscillators around the centroid.  Specifically, dispersion of the 
oscillators around their mean is governed by Eq.~(\ref{mean2}). It is also clear that the standard deviation $\sigma$ of the Hopf bifurcation (distance)
parameter $a_j$, the coupling strength $k$ and the limiting factor $\alpha$ determine the macroscopic dynamical states. 

\subsection{Bifurcation induced tipping to aging from mean-field equations}
\begin{figure*}[ht]
	\centering
	\hspace{-0.1cm}
	\resizebox{0.85\columnwidth}{!}{ \includegraphics{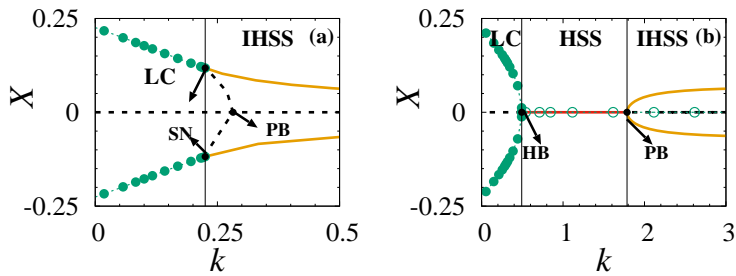} }
	\caption{(color online) Bifurcation diagrams  as a function of the coupling strength using XPPAUT for $a=-0.50$ and for two different values of the natural frequency.
	(a) $\omega= 0.1$, (b) $\omega= 0.2$.  Filled circles and solid lines correspond to stable limit-cycle oscillations and stable steady states, whereas unfilled circles and
	dashed lines correspond to unstable limit-cycle oscillations and unstable steady states.}
	\label{fig4} 
\end{figure*}

\begin{figure*}[ht]
	\centering
	\hspace{-0.1cm}
	\resizebox{0.65\columnwidth}{!}{ \includegraphics{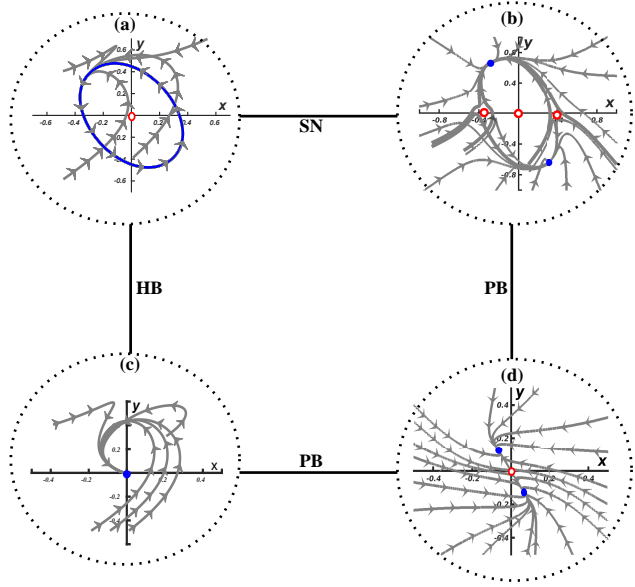} }
	\caption{(color online) Phase space trajectories near  (a) Limit-cycle oscillation, (b) IHSS  with  saddles (open circles) and nodes (filled circles),  
	(c) HSS and (d) Two stable nodes (filled circles)
	along with a unstable node at origin (open circle) near pitch-fork bifurcation.}
	\label{fig5} 
\end{figure*}

We have depicted the real part of $Z$ in the bifurcation diagrams, plotted using XPPAUT~\cite{be2002} from the evolution equations corresponding to
 the centroid $Z$, Eqs.~(\ref{mean1})  and ~(\ref{mean2}),
as a function of $k$ for two different natural frequencies in Figs.~\ref{fig4}.   Stable limit-cycle oscillations, IHSS and HSS are indicated by filled circles,  yellow/light grey solid lines
and red/dark grey solid line, respectively.  Open circles correspond to unstable limit-cycle oscillations and dashed line corresponds to the unstable steady states. 
The transition from limit-cycle oscillations (filled circles) to IHSS (indicated by yellow/light grey solid lines)  is shown in Fig.~\ref{fig4}(a) for $\omega=0.1$,
 which is mediated by the saddle-node bifurcation.  The stable limit-cycle oscillations
obtained from Eqs.~(\ref{mean1}) and ~(\ref{mean2}), just before the saddle-node bifurcation,   is depicted  in Fig.~\ref{fig5}(a), where the
 unstable steady state at the origin is indicated by an open circle. 
Trajectories starting from different initial states eventually settle on to the stable limit-cycle.
The dynamics near the saddle-node
bifurcation is depicted in Fig.~\ref{fig5}(b), where stable nodes are indicated by the filled circles and saddle points are by open circles. 
Trajectories starting from  the  initial states in the basin of the stable nodes reach  them, whereas the trajectories emerging near the saddle points slip away from
them and settle at the nearby stable nodes as depicted in  Fig.~\ref{fig5}(b).
 The saddle points collide with  each other resulting in an unstable steady state  though the pitchfork bifurcation at a large value of the coupling strength 
 away from the saddle-node bifurcation as can be observed from Fig.~\ref{fig4}(a),  and the dynamics after the pitchfork bifurcation
is indicated in Fig.~\ref{fig5}(d).
Transition from macroscopic oscillatory state  to IHSS via aging is depicted in  Fig.~\ref{fig4}(b) for $\omega=0.2$ as a function of the coupling strength. 
 The transition to the aging state (red/dark grey solid line)
 from the macroscopic oscillatory state (filled circles), that is tipping to  the aging, occurs via the Hopf bifurcation.  Similarly, the transition from the aging to the 
 IHSS (yellow/light grey solid lines) is mediated by the pitch-fork bifurcation (see  Fig.~\ref{fig4}(b)). 
{\bf The stable limit-cycle oscillation (filled circle) looses its stability and becomes unstable (open circles) at the Hopf bifurcation, which then coexists with the
aging state.  The unstable limit-cycle oscillation prevails even beyond the pitchfork bifurcation and coexists with the unstable steady state (origin) and the
IHSS.}  The dynamics near the Hopf bifurcation is shown in 
Figs.~\ref{fig5}(a)  and ~\ref{fig5}(c). All the trajectories emanating from the neighborhood of stable HSS  reaches it as depicted in Fig.~\ref{fig5}(c). The dynamics just before 
and after the pitchfork bifurcation is captured in Figs.~\ref{fig5}(c)  and ~\ref{fig5}(d), respectively, 

Now, the stability of the homogeneous steady state (aging state) can be deduced, following the Routh-Hurwitz criterion,  from the characteristic equation 
corresponding to  Eqs.~(\ref{mean1}) and ~(\ref{mean2}).  The critical  curves corresponding to the Hopf and pitch-fork bifurcations can be obtained as
\begin{align}
 f_{HB}(k,a,\omega,\sigma,\alpha)&\,=\omega ^2 \beta_1^2 ((2 a-\alpha  k) \beta_2-4 \sigma )+2 a \beta_3 \left(\beta_2^2 (\alpha  k-2 a)^2-\sigma  \beta_1^2\right)=0,
  \label{fhb}
   \end{align}
and
\begin{align}
 f_{PB}(k,a,\omega,\sigma,\alpha)&\,=\omega ^2 \left(a \beta_3+2 \sigma ^2\right)\\ \nonumber
 &\,+(a^2-\sigma^2 ) \left((a-\alpha  k) (a-\alpha  k+k)-\sigma ^2\right)+\omega ^4=0,
  \end{align}
where $\beta_1=4 a-2 \alpha  k+k$,  $\beta_2=2 a-\alpha  k+k$,  $\beta_3=2 a-2 \alpha  k+k$. For $\alpha=1$, from  the condition $f_{PB}(k,a,\omega,\sigma,\alpha)=0$, one
can explicitly deduce the critical curve corresponding to the pitch-fork bifurcation as
\begin{eqnarray}
k_{PB}=\frac{-2 a^2 \sigma ^2+2 a^2 \omega ^2+a^4+2 \sigma ^2 \omega ^2+\sigma ^4+\omega ^4}{a \left(a^2-\sigma ^2+\omega ^2\right)}
\label{kpb}
\end{eqnarray}
\begin{figure*}[ht]
	\centering
	\hspace{-0.1cm}
	\resizebox{0.85\columnwidth}{!}{ \includegraphics{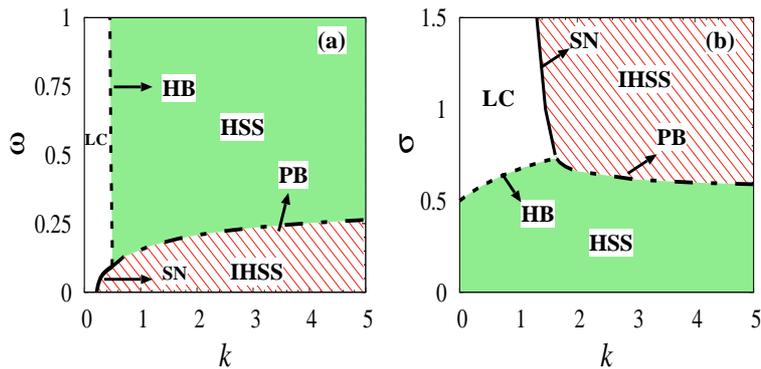} }
	\caption{(color online) Analytical two parameters phase diagrams for  $a=-0.5$  in the (a) $(k, \omega)$ space  for $\sigma$=0.60, and   (b) $(k, \sigma)$ space  for $\omega$=0.25.}
	\label{fig6} 
\end{figure*}
Analytical critical curves demarcating the dynamical regions are shown as two-parameter phase diagrams in the $(k, \omega)$ space in Fig.~\ref{fig6}(a)
and in the $(k, \sigma)$ space in Fig.~\ref{fig6}(b) for $a=-0.5, \omega=0.25$ and $\sigma=0.60$.
The critical stability curve corresponding to the pitchfork bifurcation,  Eq.~(\ref{kpb}), is depicted as a dashed-dotted line 
across which there is a tipping to the aging state.  Similarly, there is also tipping to aging from the limit-cycle oscillations via Hopf bifurcation curve (indicated by a dotted line)
obtained from  Eq.~(\ref{fhb}).  The saddle-node bifurcation curve, indicated by a solid line, mediating the transition from the macroscopic oscillatory state to IHSS 
is actually obtained from XPPAUT. The dynamical transitions and their spread in Figs.~\ref{fig6}(a)  and ~\ref{fig6}(b) fairly well agree  with that in the 
numerical figures in Figs.~\ref{fig2}(a)  and \ref{fig3}(b), respectively.
% in the limit of large coupling strength,  assumption made to deduce the mean-field equations.

\subsection{Effect of the limiting factor}
\begin{figure*}[ht]
	\centering
	\hspace{-0.1cm}
	\resizebox{0.85\columnwidth}{!}{ \includegraphics{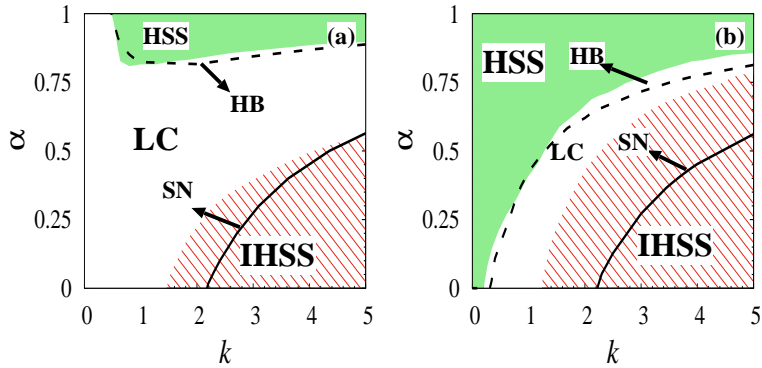} }
	\caption{(color online) Two parameters phase diagrams  in the  $(k, \alpha)$ space for (a)  $a=-0.5$, and (b) $a=-0.7$.  The other parameters are 
	$\sigma$=0.60, and $\omega$=0.25.}
	\label{fig7} 
\end{figure*}

In this section, we will unravel the effect of the limiting factor $\alpha$ on the observed dynamical regions.  Two parameter phase diagrams in 
 the $(k, \alpha)$ space for  $a=-0.5$, and  $-0.7$ are shown in Figs.~\ref{fig7}(a) and ~\ref{fig7}(b), respectively. The standard deviation is chosen as $\sigma=0.60$ 
 and the natural frequency of oscillation is fixed as $\omega=0.25$.  The analytical critical curve (Hopf bifurcation curve) is  shown as a dashed line, whereas
 the saddle-node bifurcation curve obtained from XPPAUT is depicted as a solid line. Shaded and unshaded regions are obtained by numerically solving 
 the dynamical equation (\ref{model1}) of the globally coupled Stuart-Landau limit-cycle oscillators.  
  As the limiting factor is decreased from the unity value, the spread of the aging state 
 starts to decrease first at large values of $k$ and this behavior gradually extends to smaller values of $k$.  Below a threshold value of $\alpha$, the aging
 state (HSS) completely disappears for $a=-0.5$  (see Fig. ~\ref{fig7}(a)) leading to the macroscopic oscillatory state thereby increasing the robustness 
 of the globally coupled Stuart-Landau limit-cycle oscillators.  On the other hand, a finite but narrow range of aging state persists  even at $\alpha=0$ for
 $a=-0.7$ (see Fig. ~\ref{fig7}(b)) as there exists oscillators only  with $a_j<0$. Decreasing the limiting factor  favors the spread of the IHSS  to a 
 large region of the coupling strength.  Limit-cycle oscillations  emerge in between the  homogeneous and inhomogeneous steady states. Thus it is evident
 that the limiting factor favors the macroscopic oscillatory state (see Fig. ~\ref{fig7}(a))  even in the presence of a large fraction of inactive oscillators, whereas it favors largely IHSS 
 for all inactive oscillators in addition to the macroscopic oscillatory state (see Fig. ~\ref{fig7}(b)).

\section{Summary and Conclusions}
We considered a heterogeneous network of $N$ globally coupled Stuart-Landau limit-cycle oscillators along with  the symmetry breaking coupling.
The competing interaction between the heterogeneity, which is increased by increasing the standard deviation $(\sigma)$ of the Hopf bifurcation parameter of
the Stuart-Landau oscillator about a mean, and  the symmetry breaking coupling  is explored in considerable detail.  The globally coupled oscillators
exhibit limit-cycle oscillations, homogeneous (aging) and inhomogeneous steady states.  Transitions among these dynamical states depend on 
the natural frequency of oscillations resulting in the state dependent aging. Increasing the heterogeneity is found to favor the limit-cycle oscillations
in the entire range of the natural frequency for low values of  coupling, whereas  strong symmetry breaking coupling is found to favor
the IHSS state. The threshold value of $\omega$ for the tipping to aging from the oscillatory state  is also increased for increasing heterogeneity 
of the coupled oscillator network. The aging state predominates  most  of the parameter space for the values of the standard deviation of the Hopf bifurcation parameter
below its mean value as there exist only inactive oscillators in the entire network.  The limiting factor in the diffusive coupling favors the macroscopic oscillatory state
even in the presence of a large fraction of inactive oscillators in the network thereby increasing the robustness of the network.  Surprisingly, a finite region of 
macroscopic oscillatory state is also found even when all the oscillators are inactive upon limiting the diffusive interaction.

The globally coupled  limit-cycle oscillators are reduced
to a system of two evolution equations for the macroscopic order parameters, corresponding to the mean-field and the shape parameter, using the self-consistent 
field approach in the strong coupling limit and  narrow distribution of  the heterogeneous parameter.  The bifurcation diagrams obtained from the macroscopic order parameters
elucidate different bifurcation scenarios  responsible for the dynamical transitions observed in $N$ globally coupled Stuart-Landau limit-cycle oscillators.
In particular, Hopf and pitchfork bifurcations are found to induce the tipping to the aging state.  The  dynamical states demarcated by the 
critical stability curves (bifurcation curves), deduced from
the characteristic equation corresponding to the  macroscopic order parameters,   are found to agree fairly well with the two-parameter phase diagrams obtained
by numerically solving the $N$ globally coupled Stuart-Landau limit-cycle oscillators.  The bifurcation diagrams obtained from the evolution equations of the 
macroscopic order parameters clearly elucidate the bifurcation induced tipping to the aging state in addition to the other dynamical transitions  as a result
of the competing interactions between the heterogeneity  and the symmetry breaking coupling.

\bigskip

\textbf{Acknowledgement:} IG wishes to thank SASTRA Deemed University for research fund and extending infrastructure support to carry out this work.
The work of VKC forms part of a research project sponsored by CSIR Project under Grant No. 03(1444)/18/EMRII and  SERB-DST- MATRICS Grant No. MTR/2018/000676. 
 The work of ML is
supported by the Department of Science and Technology Science and Engineering Research Board Distinguished
Fellowship (Grant No. SERB/F/6717/2017-18).

\textbf{Author contribution statement:}  VKC and DVS formulated the problem. IG carried out the simulations and calculations.  All the
authors  discussed the results and drafted the manuscript.

\end{document}